\documentclass[aps,showpacs,preprintnumbers,amsmath,amssymb,
nofootinbib,showkeys,floatfix]{revtex4}

\usepackage{latexsym}
\usepackage{amsmath}
\usepackage{amssymb}

\usepackage{supertabular} 
\usepackage{placeins}
\usepackage{epsfig}
\usepackage{graphicx}
\usepackage{graphics}

\usepackage{hyperref} 
\hypersetup{
   colorlinks=true,
   linkcolor=blue,
   citecolor=blue,
   urlcolor=blue
}

\begin{document}
\title{Quark model description of the $\Lambda_c(2940)^+$ as a molecular $D^*N$ state and the possible 
existence of the $\Lambda_b(6248)$}
\author{P.G. Ortega$^{1}$
\footnote{
Currently at CERN (European Organization for Nuclear Research), CH-1211 Geneva,
Switzerland and
Instituto de F\'isica Corpuscular, Universidad de Valencia, 46071 
Valencia, Spain},
D.R. Entem$^1$, F. Fern\'{a}ndez$^1$.}
\affiliation{
$^1$ Grupo de F\'{\i}sica Nuclear and IUFFyM,
Universidad de Salamanca, E-37008 Salamanca, Spain \\
}

\begin{abstract}
The $\Lambda_c(2940)^+$ baryon is studied in a constituent quark model as a molecular state composed by nucleons and $D^*$ mesons. 
A bound state with the right binding energy is found for the $J^\pi=\frac{3}{2}^-$ channel. The partial widths
$\Lambda_c(2940)^+\rightarrow ND$ and $\Lambda_c(2940)^+\rightarrow \Sigma_c\pi$ are calculated and the results are consistent with the experimental data.
Additionally a  bottom partner $\bar B^*N$ is predicted with a mass of 6248 MeV/$c^2$.
\end{abstract}

\maketitle

In the past years some new charmed baryons such as $\Lambda_c(2765)$, $\Lambda_c(2880)$ and $\Lambda_c^+(2940)$
were observed by Belle, BaBar and Cleo Collaborations~\cite{BB1,BLL1,CL1}.
Together with the well established $\Lambda_c^+$, $\Lambda_c^+(2595)$ and $\Lambda_c(2625)$ complete the charmed lambda spectrum until now.

The $\Lambda_c(2940)^+$ was originally observed by BaBar~\cite{BB1} analyzing the $D^0p$ invariant mass spectrum. 
The measured mass and width were respectively $M=2939,8\pm 1.3\pm 1.0$ MeV/$c^2$
and $\Gamma=17.5\pm 5.2\pm 5.9$ MeV. It was later confirmed by Belle~\cite{BLL1} as a resonant structure in the 
$\Sigma_c(2455)\pi$ decay, with a mass of $2938.0\pm 13^{+2.0}_{-4.0}$ MeV/$c^2$ and a width of 
$\Gamma=13^{+ 8}_{-5}\,^{+27}_{-7}$ MeV compatible with BaBar measurement. Belle also reported in the same reference an important result to understand the nature of the $\Lambda_c(2940)^+$, mainly setting the spin and parity of the $\Lambda_c(2880)^+$ to be $5/2^+$.

The $\Lambda_c$ spectrum has been studied, among others, in relativized~\cite{ISG1} or relativistic~\cite{EBR1} quark models. 
Both models are able to reproduce the mass of the $\Lambda_c(2880)^+$ as a $5/2^+$ state. Concerning the 
$\Lambda_c(2940)^+$, in Ref.~\cite{ISG1} a 
$5/2^-$ state is found in the $2900$ MeV/$c^2$ mass region but this state cannot be identified with the $\Lambda_c(2940)^+$ 
because the experimentally observed $\Lambda_c(2940)^+\rightarrow D^0 p$ is forbidden for this quantum numbers. 
No candidates around $2900$ MeV/$c^2$ are found in Ref.~\cite{EBR1}. Therefore these authors suggested that the so called 
$\Lambda_c(2940)^+$ may be the first radial (2$S$) excitation of the $\Sigma_c$ with $J^P=\frac{3}{2}^+$ quantum numbers. 
However this assignment is not compatible with the fact that there is no evidence in the $D^{+}p$ of doubly charged partners.

A notable observation is that the mass of the $\Lambda_c(2940)^+$ is just $6$ MeV/$c^2$ below the $D^{*0}p$ threshold. 
Therefore is very tempting to look at it as a  $D^{*0}p$ molecular state.

This possibility has been first studied by He {\it et al.}~\cite{HE}. Using an effective Lagrangian to describe the coupling 
of the $\Lambda_c(2940)^+$ with its constituents they found that the molecular structure can explain the $\Lambda_c(2940)^+$ 
as a $D^{*0}p$ $J^P=\frac{1}{2}^-$ state. However the authors were not able to calculate the absolute value of the decay width.

Using also an effective Lagrangian approach, Dong {\it et al.}~\cite{DON} have studied the $\Lambda_c(2940)^+$  as a 
$J^P=\frac{1}{2}^+$ or $J^P=\frac{1}{2}^-$ molecular state. The $\Lambda_c(2940)^+$ is supposed to be described by a 
superposition of $D^{*0}p$ and $D^{*+}n$ components. In their results the possibility of the $J^P=\frac{1}{2}^-$ quantum 
numbers is completely excluded because the calculated partial widths are of the order of GeV. For the 
$J^P=\frac{1}{2}^+$ the decay widths are one order of magnitude lower than the experimental one being 
the dominant channels the $\Sigma_c^{++} \pi^-$ and $\Sigma_c^{0} \pi^+$ relative to the $pD^0$, contrarily to the naive expectation.
In Ref.~\cite{oset} in a unitarized coupled channel calculation the authors found a possible candidate
for the $\Lambda_c(2940)^+$ in the $3/2^-$ channel although they obtained a state with a very small
width due to the fact that they only include two meson states in $S$-waves.

In this paper we propose a theoretical explanation of the $\Lambda_c(2940)^+$ as a molecular state 
in a constituent quark model that has been extensively used to describe hadron phenomenology~\cite{US1,US2,US3}.

The model is based on the assumption that the spontaneous chiral symmetry breaking  generates the constituent mass. 
To compensate this mass term in the Hamiltonian the Lagrangian must include Goldstone boson fields which 
mediates the interaction between quarks. The minimal realization of this mechanism include one pseudoscalar and one scalar boson.
In this model baryons (mesons) are described as clusters of three quark (one pair of quark and antiquark).

The first candidates to a description of the $\Lambda_c(2940)^+$ as a $D^{*0}p$ molecule are the $J^P=\frac{1}{2}^-$
and $J^P=\frac{3}{2}^-$ states because in this case the $D^{*}$ and the $N$ are in a relative $L=0$ angular momentum. Moreover 
the most probable configuration is the $J^P=\frac{3}{2}^-$ for two reasons. First the $J^P=\frac{3}{2}^-$ has a 
sizable attractive contribution from the pseudoscalar Goldstone boson tensor component. Second, in the $J^P=\frac{1}{2}^-$ 
the pseudoscalar Goldstone boson has different sign that the scalar one whereas these two components are attractive in the 
$J^P=\frac{3}{2}^-$ state. These facts enhance the attraction in the $J^P=\frac{3}{2}^-$ state with respect to the 
$J^P=\frac{1}{2}^-$ one and increases the possibility of molecular binding.
On the other hand states with positive parity are unfavored due to the $P$-wave nature of the molecule.

As stated above, the constituent quark model used in this work has been extensively
described elsewhere~\cite{US2,US3} and therefore 
we will only summarize here its most
relevant aspects. The model is based on the assumption that the light 
constituent quark mass appears as a consequence of the spontaneous breaking of 
the chiral symmetry at some momentum scale. As a consequence the quark 
propagator gets modified and quarks acquire a dynamical momentum dependent 
mass. The simplest Lagrangian 
must therefore contain chiral fields to compensate the mass term and can be expressed as 
\begin{equation}\label{lagrangian}
{\mathcal L}
=\overline{\psi }(i\, {\slash\!\!\! \partial} -M(q^{2})U^{\gamma_{5}})\,\psi 
\end{equation}
where $U^{\gamma _{5}}=\exp (i\pi ^{a}\lambda ^{a}\gamma _{5}/f_{\pi })$,
$\pi ^{a}$ denotes the pseudoscalar fields $(\vec{\pi }
,K_{i},\eta _{8})$ with $i=1,\ldots,4$ and $M(q^2)$ is the constituent mass.
This constituent quark mass, which vanishes at large momenta and is frozen at low 
momenta at a value around 300 MeV, can be explicitly obtained from the theory but
its theoretical behavior can be simulated by parameterizing 
$M(q^{2})=m_{q}F(q^{2})$ where $m_{q}\simeq $ 300 MeV/$c^2$, and
\begin{equation}
F(q^{2})=\left[ \frac{{\Lambda}^{2}}{\Lambda ^{2}+q^{2}}
\right] ^{\frac{1}{2}} \, .
\end{equation} 
The cut-off $\Lambda$ fixes the
chiral symmetry breaking scale.

The Goldstone boson field matrix $U^{\gamma _{5}}$ can be expanded in terms of boson fields,
\begin{equation}
U^{\gamma _{5}}=1+\frac{i}{f_{\pi }}\gamma ^{5}\lambda ^{a}\pi ^{a}-\frac{1}{%
2f_{\pi }^{2}}\pi ^{a}\pi ^{a}+...
\end{equation}
The first term of the expansion generates the constituent quark mass while the
second gives rise to a one-boson exchange interaction between quarks. The
main contribution of the third term goes into two-pion exchanges which
has been simulated by means of a scalar exchange potential.

In the heavy quark sector 
chiral symmetry is explicitly broken and this type of interaction does not act. 
However it constrains the model parameters through the light meson phenomenology 
and provides a natural way to incorporate the pion exchange interaction in the 
$D^*N$ dynamics. 

Beyond the chiral symmetry breaking scale one 
expects the dynamics to be governed by QCD perturbative effects.
They are taken 
into account through the one gluon-exchange interaction derived from 
the Lagrangian 
\begin{equation}
\label{Lg}
{\mathcal L}_{gqq}=
i{\sqrt{4\pi\alpha _{s}}}\, \overline{\psi }\gamma _{\mu }G^{\mu
}_c \lambda _{c}\psi  \, ,
\end{equation}
where $\lambda _{c}$ are the SU(3) color generators and $G^{\mu }_c$ is the
gluon field. 

The other QCD nonperturbative effect corresponds to confinement,
which prevents from having colored hadrons.
Such a term can be physically interpreted in a picture in which
the quark and the antiquark are linked by a one-dimensional color flux-tube.
The spontaneous creation of light-quark pairs may
give rise at some scale to a breakup of the color flux-tube. This can be translated
into a screened potential in such a way that the potential
saturates at some interquark distance. The potential is given by
\begin{equation}
V_{CON}(\vec{r}_{ij})=\{-a_{c}\,(1-e^{-\mu_c\,r_{ij}})+ \Delta\}(\vec{%
\lambda^c}_{i}\cdot \vec{ \lambda^c}_{j})\,
\end{equation}
where $\Delta$ is a global constant to fit the origin of
energies. Explicit 
expressions for all these interactions are given in Ref.~\cite{US2}.

All the parameters of the model are taken from Refs.~\cite{US1,US3} and therefore we will be able to decide 
if a molecular configuration exist or not in a very clear way.

To derive the meson baryon interaction from the $q\bar q$ interaction we use the Resonating Group Method (RGM). In our case they include a direct potential 
\begin{equation}
V_D=\sum_{i\in A;j\in B}\int \Psi^*_{l_A'm_A'}(\vec{p}_A')\Psi^*_{l_B'm_B'}(\vec{p}_B')V^D_{ij}(\vec{P}',\vec{P})\Psi_{l_Am_A}(\vec{p}_A)\Psi_{l_Bm_B}(\vec{p}_B)
\end{equation}
and an exchange one 
\begin{equation}
V_E=\sum_{i\in A,j\in B}\int \Psi^*_{l_A'm_A'}(\vec{p}_A')\Psi^*_{l_B'm_B'}(\vec{p}_B')V^E_{ij}(\vec{P}',\vec{P})\Psi_{l_Am_A}(\vec{p}_A)\Psi_{l_Bm_B}(\vec{p}_B)
\end{equation}
which gives the coupling between the different channels $D^*N$ and $\Sigma_c\pi$, done by simple quark rearrangement driven by the $q\bar q$ interaction.

To find the quark-antiquark bound states we solve the 
Schr\"odinger equation 
using the Gaussian Expansion Method~\cite{r20}. 
For the baryon wave function we use a gaussian form with a suitable parameter,
\begin{equation} 
\psi (\vec{p}_i)=\prod_{i=1}^3 \left[ \frac{b^2}{\pi}\right]^{\frac{3}{4}} e^{-\frac{b^2 p_i^2}{2}}
\end{equation}
where $b=0.518$ fm is taken from Ref~\cite{US1} . In terms of Jacobi coordinates this wave function is expressed as,
\begin{equation} 
\psi =\left[ \frac{b^2}{3\pi}\right]^{\frac{3}{4}}e^{-\frac{b^2P^2}{6}} \phi _B (\vec{p}_{\xi_1},\vec{p}_{\xi_2})
\end{equation}
where $\vec{P}$ is the baryon momentum in the center of mass system and $\vec{p}_{\xi_1}$ and $\vec{p}_{\xi_2}$ momenta correspond to internal coordinates. The internal spatial wave function is written as,
\begin{equation} 
\phi _B (\vec{p}_{\xi_1},\vec{p}_{\xi_2})=\left[\frac{2b^2}{\pi}\right]^{\frac{3}{4}}e^{-b^2p^2_{\xi_1}}\left[\frac{3 b^2}{2\pi}\right]^{\frac{3}{4}}e^{-\frac{3}{4} b^2p^2_{\xi_2}}
\end{equation}
%


Exploiting the symmetries of the system there are four possible diagrams which contribute to rearrangement processes.
The $h_{fi}$ matrix elements corresponding to each diagram is the product of 
three factors
\begin{eqnarray} 
	h_{ij}(\vec P',\vec P) &= S& 
	\langle \phi_{\Sigma} \phi_{\pi}|H_{ij}^O| \phi_{D^*} \phi_{N} \rangle
	\langle \xi_{\Sigma\pi}^{SFC} |\mathcal{O}_{ij}^{SFC}| 
\xi_{D^*N}^{SFC} \rangle
\end{eqnarray}
where $S$ is a phase characteristic of each diagram, resulting from the 
permutation between fermion operators. The orbital part involves contributions 
of both the One-Gluon-Exchange $V_{OGE}$ and Confinement $V_{CON}$ potential 
and can be written as [e.g. for the case $(ij)=(a\bar b)$]
\begin{eqnarray}
	\langle \phi_{\Sigma} \phi_{\pi}|H_{ij}^O| \phi_{D^*} \phi_{N} \rangle 
&=&
	\int d^3 P_{\Sigma}d^3 P_{\pi}d^3 P_{D^*}d^3 P_{N} \,\,
	\phi^*_{\Sigma}(P_{\Sigma}) \phi^*_{\pi}(P_{\pi}) 
	\delta(\vec{P}_{\pi}-\vec{P}_{D^*})
	\nonumber \\ &&
	\delta(\vec{P}_{\pi}-\vec{P}_{N}-(\vec{P}'-\vec{P}))
	H(-\frac 1 2(\vec{P}_{D^*}+\vec{P}_{N})+\vec{P}_{\Sigma}
	+\frac 1 2(\vec{P}'-\vec{P}))
	\nonumber \\ &&
	\phi_{D^*}(P_{D^*}) \phi_{N}(P_{N}).
\end{eqnarray}

The coupled channel equations are solved through the Lippmann-Schwinger equation for the t matrix
\begin{equation}
 t^{\beta\beta'}( p, p',E)=V^{\beta\beta'}_T( p, p',E)-\sum_{\beta''}\int dq q^2 \frac{V^{\beta\beta''}_T( p, q,E)t^{\beta''\beta'}( q, p',E)}{q^2/(2\mu)-E-i0}
\end{equation}
where $\beta$ specifies the quantum numbers necessary to define a partial wave in the baryon meson state.
Finding the poles of the $t(\vec p,\vec p',E)$ matrix we will determine the mass of the possible molecule.

The decay of the particle is calculated through the standard formula:
\begin{equation} \label{ec:decayL2940}
 \Gamma = 2\pi \frac{E_AE_Bk_0}{M_\Lambda} \sum_{J,L}|\mathcal{M}_{J,L}|^2
\end{equation}
where $E_A$ and $E_B$ are the relativistic energies of the final state hadrons $DN$ or $\Sigma_c\pi$, $M_\Lambda$ is the mass of the molecule and $k_0$ is the on-shell momentum of the system, given by,
\begin{equation}
 k_0=\frac{\sqrt{[M_\Lambda^2-(M_A-M_B)^2][M_\Lambda^2-(M_A+M_B)^2]}}{2M_\Lambda}
\end{equation}

If we assume a $D^*N$ structure for the $\Lambda_c$, in order to obtain the decay we must calculate the amplitude to the final state in the constituent quark model framework.
For $DN$ state this can be achieved with a direct potential, but the $\Sigma_c\pi$ state requires a rearrangement diagram, as stated before.

To calculate the final amplitude of the process $\mathcal{M}$ the wave function of the molecular state is used,
\begin{equation} \label{ec:amplitudL2940}
 \mathcal{M}=\int_0^\infty V_{D^*N\to AB}(k_0,P)\chi_{D^*N}(P)\,P^2dP
\end{equation}
where $V_{D^*N\to AB}(k_0,P)$ is the potential to the final state and $\chi_{D^*N}$ is the molecule wave function.


We have studied the two negative parity states $J^P=\frac{1}{2}^-$ and  $J^P=\frac{3}{2}^-$. We found a molecular state with 
$J^P=\frac{3}{2}^-$ quantum numbers with a mass of $2938.68$ MeV/$c^2$. One can see from Table~\ref{t1} that the molecule is 
basically a $^4S_{3/2}$ $D^*N$ state with a small mixture of  $D_{3/2}$ states. 

\begin{table}
\begin{center}
\begin{tabular}{c|ccc|cc|cc}
\hline
\hline
$M\,(MeV)$ & $\mathcal P_{^4S_{3/2}}$ & $\mathcal P_{^2D_{3/2}}$& $\mathcal P_{^4D_{3/2}}$
& $\mathcal P_{{D^*}^0p}$ & $\mathcal P_{{D^*}^+n}$&$\mathcal P_{I=0}$ & $\mathcal P_{I=1}$ \\
\hline
   2938,80& 96,22 & 0,86 & 2,92& 63,93  & 36,07 & 97,52 & 2,48 \\
 
\hline
\hline
\end{tabular}
\caption{\label{t1} Mass of the $\Lambda_c^+(2940)$ and different channel probabilities.}
\end{center}
\end{table}

\begin{table}
\begin{center}
\begin{tabular}{c|c|c|c}
\hline
\hline
Decay channel & Width (MeV) & decay channel& Width (keV) \\
\hline
$\Lambda_c^+ \rightarrow D^0p$ & 9.42 & $\Lambda_c^+\rightarrow \Sigma_c^{++} \pi^-$ & 29.7 \\
$ \Lambda_c^+ \rightarrow D^+n$ & 10.74 & $\Lambda_c^+\rightarrow \Sigma_c^{+} \pi^0$ & 25.2 \\
&  & $\Lambda_c^+\rightarrow \Sigma_c^0 \pi^+$ & 21.1 \\
\hline
\hline  
 $\Gamma (total)$ & 20.2 & $\Gamma (experimental)$ & $17^{+8}_{-6}$ \\
\hline
\hline
\end{tabular}
\caption{\label{t2} Widths of the $\Lambda_c^+(2940)$ for different decay channels.}
\end{center}
\end{table}

If one looks to the different components in the charge basis one sees that the $\Lambda_c(2940)^+$ is  a $D^{0*}p$ molecule with a sizable  
$D^{*+}n$ component. Looking to the isospin, one realizes that the state is almost a pure $I=0$ state as required by the experimental data.

We have study the decay channels $\Lambda_c(2940)^+\rightarrow D^0p$, $\Lambda_c(2940)^+\rightarrow D^+n$,
$\Lambda_c(2940)^+\rightarrow \Sigma_c^{++} \pi^-$, $\Lambda_c(2940)^+\rightarrow \Sigma_c^{+} \pi^0$ and
$\Lambda_c(2940)^+\rightarrow \Sigma_c^0 \pi^+$.
One can expect a larger decay width to the $DN$ channels because the decay to the $\Sigma_c \pi$ channels occurs 
through a rearrangement process and so the coupling of these channels with the molecule is weaker.

The results for the different decay widths are shown in Table~\ref{t2}.
The  widths of the $D^0p$ and $D^+n$ decay channels are of the order of the experimental data whereas
the contributions of the $\Sigma_c\pi$ channels are negligible. The predicted total width  agrees with experimental data.

Having obtained a molecule in the $D^*N$ channel one can expect to find a similar structure in the bottom sector. 
In fact, in this sector the quark kinetic energy is lower which will favor the cluster formation.

We have redone the calculation with the only change of the heavy quark mass obtaining a $\bar B^*N$ molecule on the 
$J^P=\frac{3}{2}^-$ channel with a larger binding energy with respect to its charmed partner. The calculated mass is 
$M=$6248 MeV/$c^2$. In Table~\ref{t3} we show the probabilities for the different molecular components. 
As before we obtain a $I=0$ molecule with similar $B^{*-}p$ and $\bar B^{*0}n$ components. 

The decay widths through the different channels are shown in Table~\ref{t4}. In the case of the 
$B^-p$ and $\bar B^0n$ channels we obtain lower values than in the charm sector due to the less available phase space. 
The values of the decay widths through the $\Sigma_b \pi$ channels are, as in the former case, two orders of magnitude lower.  

\begin{table}
\begin{center}
\begin{tabular}{c|ccc|cc|cc}
\hline
\hline
$M\,(MeV)$ & $\mathcal P_{^4S_{3/2}}$ & $\mathcal P_{^2D_{3/2}}$& $\mathcal P_{^4D_{3/2}}$
& $\mathcal P_{{B^*}^-p}$ & $\mathcal P_{{\bar B}^{*0} n}$&$\mathcal P_{I=0}$ & $\mathcal P_{I=1}$ \\
\hline
   6248.34& 95.15 & 1.08 & 3.77& 52.56  & 47.44 & 99.91 & 0.09 \\
 
\hline
\hline
\end{tabular}
\caption{\label{t3} Mass of the $\Lambda_b(6248)$ and different channel probabilities.}
\end{center}
\end{table}

\begin{table}
\begin{center}
\begin{tabular}{c|c|c|c}
\hline
\hline
Decay channel & Width (MeV) & Decay channel& Width (keV) \\
\hline
$ \Lambda_b \rightarrow B^-p $ & 3.69 & $\Lambda_b\rightarrow \Sigma_b^{+} \pi^- $ & 40.9 \\
$\Lambda_b \rightarrow \bar B^0n $ & 3.75 & $\Lambda_b\rightarrow \Sigma_b^{0} \pi^0 $ & 39.5 \\
&  & $\Lambda_b\rightarrow \Sigma_b^- \pi^+$ & 38.1 \\
\hline
\hline
\end{tabular}
\caption{\label{t4} Widths of the $\Lambda_b(6248)$ for different decay channels.}
\end{center}
\end{table}

As a summary we have calculated the $\Lambda_c(2940)^+$ baryon as a $D^*N$ molecule in a constituent quark model. 
We obtain a molecule in the $J^P=\frac{3}{2}^-$ channel and $I=0$ with a mass and total width which agree with the experimental data. 
Moreover we predict in the same model the existence of a new resonance, the $\Lambda_b(6248)$, with a relatively small width 
$\Gamma= 7.5$ MeV which can be discovered at LHCb in the near future.

\acknowledgments
This work has been partially funded by Ministerio de Ciencia y Tecnolog\'ia
under Contract No. FPA2010-21750-C02-02, by the European Community-Research
Infrastructure Integrating Activity 'Study of Strongly Interacting Matter'
(HadronPhysics3 Grant No. 283286), the Spanish Ingenio-Consolider 2010 Program
CPAN (CSD2007-00042).

\end{document}